\documentclass[10pt,a4paper,twocolumn]{article}
\usepackage[utf8]{inputenc}
\usepackage{amsmath}
\usepackage{amsfonts}
\usepackage{amssymb}
\usepackage{graphicx}
\usepackage{yfonts}
\usepackage{xargs}
\usepackage{url}
\usepackage[pdftex,dvipsnames]{xcolor}  
\usepackage[noadjust]{cite}  
\usepackage{changes} 
\usepackage{algorithm}
\usepackage{algpseudocode}
\usepackage[font={small,it}]{caption}
\usepackage{tikz}
\usepackage{textcomp}
\usepackage{array,makecell,multirow,tabularx,ragged2e,booktabs}
\usepackage[font=small,skip=5pt]{caption}
\usepackage[a4paper, total={6.9in, 9in}]{geometry}

\usepackage[colorinlistoftodos,prependcaption,textsize=tiny]{todonotes}
\newcommandx{\change}[2][1=]{\todo[linecolor=blue,backgroundcolor=blue!25,bordercolor=blue,#1]{#2}}

\newcolumntype{b}{X}
\newcolumntype{s}{>{\hsize=.75\hsize}X}

\newcolumntype{P}[1]{>{\centering\arraybackslash}p{#1}}
\newcolumntype{R}[1]{>{\RaggedLeft\arraybackslash}p{#1}}

\begin{document}
\title{Multi-feature classifiers for burst detection in single EEG channels from preterm infants}
\date{}
\author{\small X.~Navarro$^{1}$\footnotemark[1]~, F. Por\'{e}e$^{2}$~, M. Kuchenbuch$^{2,3}$~, M. Chavez$^{4}$~, A. Beuch\'ee$^{2,3}$~ and G. Carrault$^{2}$\\ 
	\tiny 
	$^1$Sorbonne Universit\'{e}s, UPMC Univ Paris 06, INSERM UMRS-1158 Neurophysiologie Respiratoire Exp\'{e}rimentale et Clinique, Paris, France \\ \vspace{-5px}
    \tiny
	$^2$Universit\'e de Rennes~1, INSERM U1099, Laboratoire Traitement du Signal et de l'Image, Rennes, France \\ \vspace{-5px}
	 \tiny
	$^3$ CHU Rennes, P\^ole de p\'ediatrie m\'edico-chirurgicale et de g\'en\'etique clinique, Rennes, France\\ \vspace{-5px}
	 \tiny
	$^4$CNRS UMR7225, H\^{o}pital de la Piti\'{e} Salp\^{e}tri\`{e}re Paris, France \\ \vspace{-5px}
	 \tiny
}

\maketitle

\begin{abstract}
The study of electroencephalographic (EEG) bursts in preterm infants provides valuable information about maturation or prognostication after perinatal asphyxia. Over the last two decades, a number of works proposed algorithms to automatically detect EEG bursts in preterm infants, but they were designed for populations under 35 weeks of post menstrual age (PMA). However, as the brain activity evolves rapidly during postnatal life, these solutions might be under-performing with increasing PMA. In this work we focused on preterm infants reaching term ages (PMA $\geq$ 36 weeks) using multi-feature classification on a single EEG channel. Five EEG burst detectors relying on different machine learning approaches were compared: Logistic regression (LR), linear discriminant analysis (LDA), k-nearest neighbors (kNN), support vector machines (SVM) and thresholding (Th). Classifiers were trained by visually labeled EEG recordings from 14 very preterm infants (born after 28 weeks of gestation) with 36 -- 41 weeks PMA. The most performing classifiers reached about 95\% accuracy (kNN, SVM and LR) whereas Th obtained 84\%. Compared to human-automatic agreements, LR provided the highest scores (Cohen's kappa = 0.71) and the best computational efficiency using only three EEG features. Applying this classifier in a test database of 21 infants $\geq$ 36 weeks PMA, we show that long EEG bursts and short inter-bust periods are characteristic of infants with the highest PMA and weights. In view of these results, LR-based burst detection could be a suitable tool to study maturation in monitoring or portable devices using a single EEG channel. 
\end{abstract}

\footnotetext[1]{Author for correspondence: xavier.navarro@upmc.fr}

\section{Introduction}
The electroencephalographic (EEG) activity in preterm infants is characterized by discontinuous patterns, alternating quiescence periods with slow, high voltage transients or bursts, continuously evolving during infancy. Inter-burst intervals (IBIs) provide valuable information about progostation and maturation as they progressively decrease in duration with increasing age in healthy preterm infants \cite{Hahn1989, Hayakawa2001}.
As preterm infants reach term age (37 to 40 PMA) EEG becomes more complex, but immature patterns are still  present \cite{Scher1992, Andre2010}. At this stage, a \emph{trac\'{e} alternant} (an alternating pattern of bursts and relatively quiet periods \cite{Lamblin1999}) predominates in quiet sleep, but long IBIs may appear, suggesting an abnormal neurodevelopmental outcome \cite{Rando2006}. Therefore, studying IBIs and transients in the EEG after term equivalent ages can be useful as an early prognostic tool \cite{Hayashi2012}.

Nevertheless, the study of IBIs is not a clinical routine in neonatal intensive care units (NICUs) and, to our knowledge, it is not existent in portable and home monitoring devices. In effect, NICUs prioritize other vital signs over electroencephalography, which is tedious and time-consuming. Moreover, the manual recognition of IBIs and bursts is, still, another laborious and subjective task requiring trained neurologists. The automation of this procedure would therefore save time costs, avoid disagreement between different annotators and, in turn, gain attractiveness as a monitoring tool in the NICU.
This challenge has motivated a number of works that propose different approaches, including supervised learning (single or multi-feature based) and clustering. The choice of the most appropriate solution is not always straightforward and relies on both clinical (e.g. infants age, developmental problems) and technical criteria (e.g number of available channels). 

The vast majority of existing algorithms have been designed for burst suppression applications both in adults (anesthesia or coma monitoring) and full-term newborns (EEG monitoring after perinatal asphyxia), but only a few works design burst detectors for preterm EEG. Even if similarities exist, burst suppression patterns are related to a clinical condition whereas preterm's bursts describe the normal EEG and evolve during postnatal life. 
In Table 1, we provide a summary of the works published over the last two decades related to preterm burst detection as well as some relevant solutions for burst suppression in full-term infants and adults. 

\begin{table*}[h!]
\footnotesize
{\def\arraystretch{1.3}\tabcolsep=6pt
\begin{tabular}{p{4.5cm} P{2.2cm} P{0.8cm} P{0.5cm} P{2.6cm} P{1.2cm} R{2.4cm}}
\toprule
\textbf{Description}	& \textbf{Patients} & \textbf{\#Chan} & \textbf{\#Sc}  & \textbf{Features} 	& \textbf{Classifier} & \textbf{Performance} \\\midrule
\hline 
 Detection of trac\'{e} alternant during sleep \cite{Turnbull2001} & 6 full-term &  14 		& 1 			& Discrete wavelet transform & S; Th & n/a  \\

Burst suppression during anesthesia \cite{Sarkela2002} & 17 adults & 1 		   & 1		 & 1 (Nonlinear energy operator)		& S; Th & Acc=94\%   \\

 EEG bursts \& heart beat ratio relationship \cite{Pfurtscheller2005} & 15 full-term & 1 & 1 & 1 (Instant. variance) & U; Th & n/a  \\

Burst suppression detection after asphyxia \cite{Lofhede2008} & 6 full-term   & 8 & 1  & 5 (Energy and frequency based)  & S; SVM & AUC=0.96  \\

Burst detection in extremely preterm \cite{Palmu2010a} & 18 preterm (23-28/28-30 w. PMA) & 1 & 2 & 1 (Nonlinear energy operator) & S; Th & Acc=90/81\%  \\

Burst suppression detection \cite{Lofhede2010} & 26 full-term & 8 & 1 & 9 (Energy and frequency based) & S; FLD & Acc=94\%  \\ 

Burst, IBI and continuous EEG detection \cite{Jennekens2011}& 8 early preterm (29-34 w PMA) & 18 & 2 & 1 (EEG power in multiple channels) & S; Th & Sn=90\% (Bursts) Sn=80\% (IBIs)  \\ 

IBI adaptive segmentation in encephalopathy \cite{Matic2012} & 8 full-term & 13 & 1 & 1 (Amplitude) & S; Th & n/a  \\

Burst detection in preterms \cite{Koolen2013} & 13 preterm (26-34 w PMA)& 9 & 2 & 1 (Line length) & S; Th & Acc=84\%   \\

Burst detection \& diagnostic interface \cite{Chauvet2014} & 394 preterm ($<$35 w PMA) & 8 & 1 &1 (Line length) & U; Clu & n/a \\ 

Burst detection in neonatal EEG \cite{OToole2014} & 10 preterm + 10 full-term & 1 & n/a &1 (Envelope derivative operator) & S; T & AUC$\ge$0.9 \\ 

EEG differentiation after asphyxia \cite{Matic2014}& 34 full-term & 12 & 1 & 3 (Amplitude and time based) & S; SVM  & Acc=84\%  \\ 

Automated detection of bursts and IBIs \cite{Murphy2015}& 36 preterm ($<$30 w GA)  & 8 & 3 & 1 (Nonlinear energy operator)& S; Th  & Algorithm/Rater: Acc=81\%, $\kappa$=0.63 ; Inter-rater: Acc=71\% $\kappa$=0.58\\

Burst/IBI classification by age \cite{Navakatikyan2016} & 26 extremely preterm  & 2 & 0 & 1 (Range EEG) & U; Th & n/a  \\ [5pt]
\bottomrule \\[-2pt]
\end{tabular} 
}
\caption{Summarized review of burst detection methods in preterm infants and other populations. Abbreviations not defined in the body text are: \#Chan, number of channels; \#Sc, number of scorers; w, weeks; S, supervised; U, unsupervised; FLD, Fisher linear discriminant; Clu, clustering; n/a not available; Acc, accuracy; Sn, sensitivity; AUC, area under receiver operating characteristic (ROC) curve; $\kappa$, Cohen's kappa.}
\normalsize
\end{table*}

If the brain activity needs to be segmented into burst and IBIs, binary classification or regression can be employed. Although this constitutes the norm of burst classification, in certain cases a third class (artefacts \cite{Sarkela2002}, continuous pattern \cite{Jennekens2011}) or other categories (classification of different degrees of activity after asphyxia \cite{Matic2012, Matic2014}) can be considered. 
  
In burst detection, single-feature detectors are often preferred when the EEG patterns are predominantly dichotomous, with low frequency deflections from the baseline that allow the use of direct measures (such as voltage amplitude) or  functions applied on the EEG (such as energy) as only feature. Dichotomous patterns can be found in a variety of altered states of consciousness \cite{Brenner1985}, but in healthy preterm infants they are characteristic of ages below 32 weeks PMA \cite{Biagioni1994}. Single-feature detectors are fast and can be easily implemented by simple thresholding, successfully employed for burst detection using a single EEG channel in very preterm infants \cite{Palmu2010} with Teager-Kaiser operator \cite{Kaiser1990}. This operator (also known as nonlinear energy operator, NLEO), and its variants \cite{Agarwal1999,OToole2014} are widely employed for EEG burst detection (see Table 1).  
For multichannel data, thresholding has been successfully applied in successive steps using EEG power \cite{Jennekens2011}, NLEO \cite{Murphy2015} or using line length as feature \cite{Koolen2013}. 

In general, using a single feature performs fairly well for very immature patterns, but as EEG complexity increases, supplementary descriptors are needed. Thus, detecting EEG patterns in full-term newborns often requires the use of several features, as those derived from wavelet analysis for \emph{trac\'{e} alternant} detection \cite{Turnbull2001} or a variety of time and frequency-based descriptors for burst suppression detectors \cite{Lofhede2008, Lofhede2010}. 

In this work, we address the detection of bursts in very preterm infants who reached term-equivalent ages (TEA). Considering the growing field of portable EEG headsets and wearable sensors, we studied the viability of burst detection using a single EEG channel. Logistic regression (LR) is evaluated by using experts visual marks and compared to other popular multi-feature methods such as linear discriminant analysis (LDA), SVM, k-nearest neighbor (kNN) and the most commonly used single-feature classifier: Thresholding (Th).

The remainder of the paper is as follows: Section \ref{data} describes the database, the evaluation subset and the construction of reference labels. In  Section \ref{meth}, we present the employed classifiers and the evaluation methodology. Section \ref{resultat} compares automatic and visual detections and shows the results of applying the LR based classifier to assess the maturation in our cohort. Finally, some concluding remarks are drawn in Section \ref{DC}.

\section{Database}
\label{data}
\subsection{EEG recordings}
Thirty-one very preterm infants, born after 27 to 29 weeks of gestation, were recorded at the CHU Hospital at Rennes (France) to study the effects of immunization (see \cite{Mialet2013} for more details about the protocol). Only pre-immunization recordings were considered in the present work to avoid eventual perturbations following the administration of vaccine. 
Infants, who presented a normal outcome and had discharged home, accounted for at least seven weeks of postnatal life (36 to 41 weeks PMA) during the recordings. The study was approved by the local institutional ethics committee (\emph{Comit\'e de Protection des Personnes}, CPP Ouest 6-598, France) and a written informed consent was given by parents. 

For each newborn, two EEG channels were acquired at sampling frequency $F_s$=512 Hz using a Brainz$^\copyright$ ~bedside monitor (Natus Medical Incorporated, San Carlos, USA) during 2 to 3 hours. Hydrogel surface electrodes were placed in fronto-parietal and temporal positions, corresponding  approximately to the  Fp1, Fp2, T3 and T4 locations of 10-20 standard systems. A bipolar reference was applied to obtain the channel pairs Fp1-T3 and Fp2-T4. Additionally, electrocardiogram (ECG), respiratory activity and hypnograms (annotations of the sleep stages in conformity to the neonatal standard \cite{Prechtl1974}) were available.

\subsection{Evaluation subset}
Labels to validate the classifiers and compute human/automated agreements was provided by two experienced neonatologists who marked manually (in burst/IBI periods) an evaluation subset.

Visual evaluations were carried out on a selection of $N_e$=14 infants (36.1 to 39.7 weeks PMA) from the above described database. Each infant provided an EEG excerpt of $D=300$ seconds. To ensure the existence of discontinuous or semi-discontinuous patterns, only segments free of artifacts in quiet sleep were considered.

Using a computer program designed to this purpose, the two clinicians (A and B) marked the bursts limits in 20-second windows displaying the pre-processed EEG. They were asked to perform the visual interpretations a second time in a different day. Visually marked bursts were then converted to discrete, binary series that coded bursts with ones and IBIs with zeros. Each category was associated, respectively with $Class \; 1$ and $Class \; 0$.   
Visual evaluations yielded four binary arrays $Y_{r,i} \in \{0,1\}$ of length $L=D \times F_s$, where $r = \{A,B\}$ represents the rater's code and $i = \{1,2\}$ is the repetition number.  

\subsection{Gold Standard}
\label{subsec:gs}
The gold standard, i.e. reference data to train and test the classification algorithms, was generated by merging the experts' evaluations. We first constructed unified marks for each rater, $Y_{r}$, by including the bursts of the two replicates:
\begin{equation}\label{eq:31}
 Y_{r} = y_{r}(k) = \left\{
  \begin{array}{l l}
    0 & \quad \text{if $y_{r,1}(k) + y_{r,2}(k)  = 0$ }\\
    1 & \quad \text{otherwise}
  \end{array} \right.
\end{equation}
for $k=1, ..., L$ and $r = \{A,B\}$. 

\begin{figure}[t!] 
   \centering
   \includegraphics[trim = 8mm 1mm 0mm 7mm,clip=true,width=\linewidth]{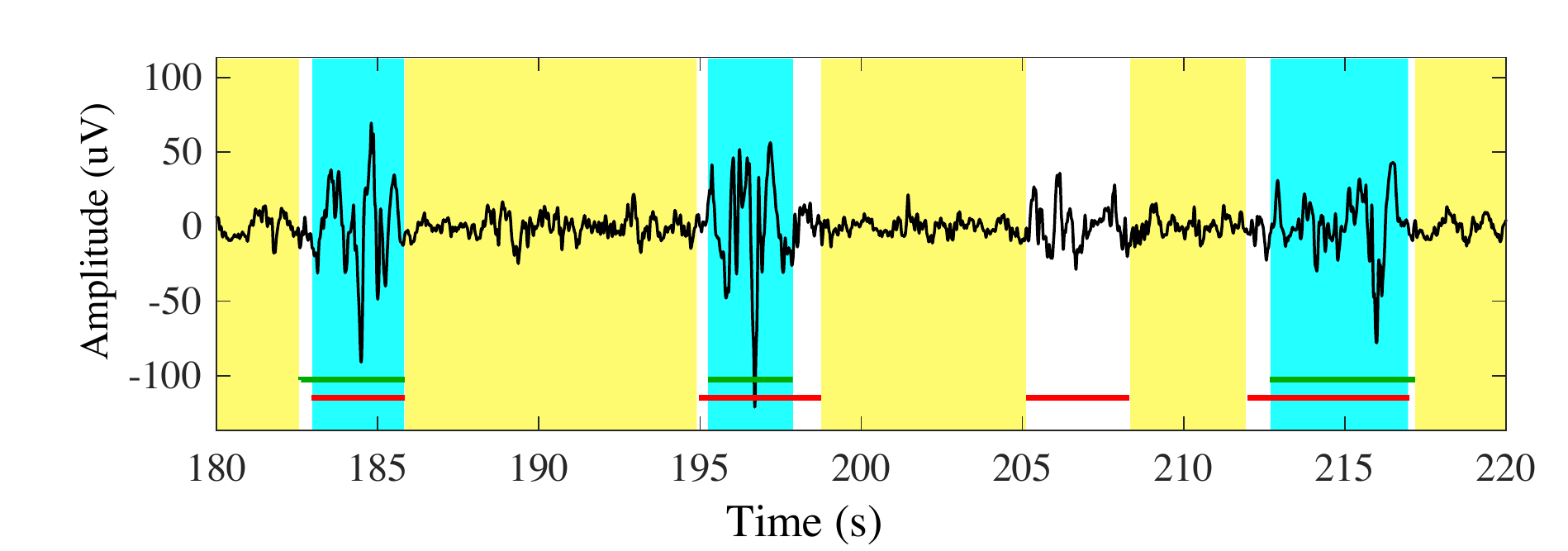} 
   \caption{Example of EEG scored with the raters labels. Green and red lines are $Y_{A}$, $Y_{B}$ respectively. Only the consensual marks are taken into account to establish the gold standard (blue areas, bursts; yellow areas, IBIs). Areas in white represent disagreeing zones.}
   \label{fig:gs}
\end{figure}

Then, we defined the gold standard, $Y$, as in \cite{Palmu2010a}, i.e. the unanimous decisions between clinicians' marks. Intervals without agreement were not considered and labeled as empty values (\o{}):
\begin{equation}
 Y = y(k) =\left\{
  \begin{array}{l l}
    1 & 	\quad \text{if $\prod\displaylimits_{r} y_{r}(k) = 1$ }\\
    0 & 	\quad \text{if $\sum\displaylimits_{r} y_{r}(k) = 0$} \\
    \mbox{\o{}} & 	\quad \text{otherwise}
  \end{array} \right.
\end{equation}
with $k=1, \ldots, L$ and $r = \{A,B\}$. An example is given in Fig. \ref{fig:gs}. 
 
\section{Methods}
\label{meth}
\subsection{Burst detection framework}
\begin{figure}[htbp] 
   \centering
   \includegraphics[trim = 0mm 2cm 0mm 1.5cm,clip=true,width=\linewidth]{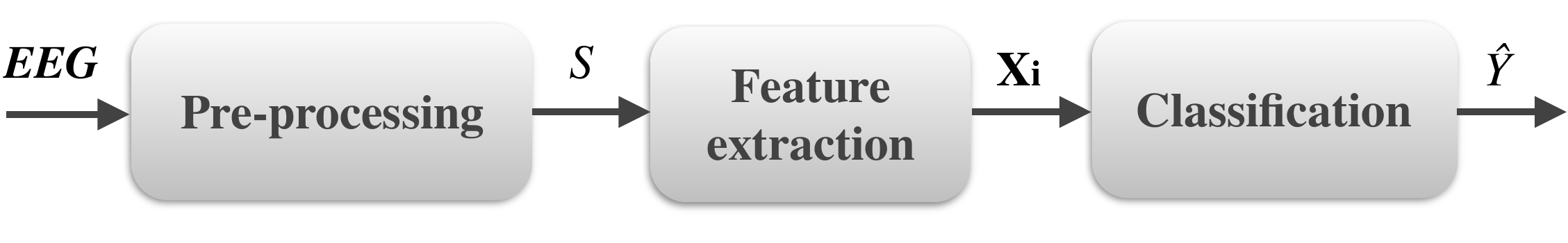} 
   \caption{Block diagram of the employed framework to detect bursts. The input signals ($EEG$) come from a two-channel system. The pre-processing block yields a one-dimensional signal, $S$, from which a feature vector, $\mathbf{X}_i$, is extracted in each window $i$. The classification block outputs a binary, one-dimensional signal, $\widehat{Y}$, with the predicted bursts.}
   \label{fig:blocks}
\end{figure}

To test the different burst detectors, we employed the general scheme that can be divided in three main blocks: pre-processing, feature extraction and classification (see Fig.~\ref{fig:blocks}). The content of some blocks depend on the classification approach.  

\subsubsection*{Pre-processing}
In this block, signal-to-noise ratios in EEG signals are improved by applying artifact correction or rejection and filters.
Due to prone position and nursing, certain artifacts are typically more abundant in one channel. Here, the less contaminated one is selected regarding its statistical properties (variance, kurtosis, joint probability of EEG activity values \cite{Delorme2007}). Then, a linear phase band-pass filter whose respective lower and upper cut-off frequencies are set to $F_l$ and $F_u$, is applied. Cut-off frequencies can vary depending if baseline and high frequency noise needs to be attenuated or if a specific bandwidth wants to be enhanced. The resulting signal, $S$, is finally sub-sampled to 128 Hz.


\subsubsection*{Feature extraction}
This block computes a number of functions on $S$ to obtain a feature vector $\mathbf{X_i} \in \mathbb{R}^{N_f}$ with $N_f$ the number of features, and $i=1, \ldots N_w$, with $N_w$ the number of windows of $S$.
Features, chosen to capture pronounced characteristics in bursts (see Table 2), have already been exploited to solve EEG classification problems. Since $\mathbf{X_i}$ are computed in overlapping $W$-second windows, the effective sampling rate of features with respect to $S$ is reduced. Hence slow trends were enhanced over fast transients by performing a smoothing of  $\mathbf{X_i}$ as suggested by \cite{Lofhede2008}. To this purpose we applied a 1-dimensional, 10th order median filter on each feature \cite{Arce2005}. Finally, to avoid outliers that might decrease classification performances, $\mathbf{X_i}$ was quantile-normalized to impose their values to fall within the 1st and 99st percentiles \cite{Lofhede2008}.

\begin{table}[t!]
\footnotesize
\begin{tabular}{p{0.6cm} p{7cm}}
\toprule
\thead{Name} & \thead{Description}  \\\midrule
\hline \\[-2pt]
$Mm$ 				& Difference between the maximum and the minimum value \\[2pt]

$DM$ 				& Maximum of absolute values of the discrete difference:
    						\begin{equation*}
    						DM = \max_{k=1,.., l} \{ |S(k) - S(k-1)|\},
  							\end{equation*} 
  							where $l$ is the number of points in $W$\\[2pt]
$SD$ 				& Standard deviation 		\\	[2pt]				
$Kt$ 				& Kurtosis \\[2pt]
$NL$             &Nonlinear energy operator (NLEO) \cite{Sarkela2002}:
  							\begin{equation*} 				
    						NL  =  \frac{1}{l} \sum_{k=1}^{l} S(k)S(k-3) - S(k-1)S(k-2) .
  						\end{equation*} \\[-2pt]
$AD$ 			  & Averaged differentiation, defined as:
    						\begin{equation*}
     						AD = \frac{1}{l} \sum_{k=1}^{l} |S(k) - S(k-1)|.
    					\end{equation*} \\[-2pt]
$Hs$ 			& Shannon Entropy \cite{Lofhede2008}:
						\begin{equation*}
							Hs = - \sum_{q} p(I_q) \log p(I_q) ,
						\end{equation*}
					where $p(I_q)$, $q=1...Q$ is a discrete set of probabilities estimated by counting the $l$ points within $Q=$16 histogram bins.  \\[2pt]
 $Pw$ 		&  Power between 0.5 -- 3Hz, estimated by an auto-regressive model using the Burg method. Model order (15) was set to the mean value provided by Akaike's information criterion \cite{Akaike1998}. \\
\bottomrule 					 											
\end{tabular}
\caption{Definition of the features applied on each EEG window ($W$ seconds) for multi-feature classifiers.}
\end{table}

\subsubsection*{Classification}
The purpose of this block is to identify the labels from new observations using $\mathbf{X_i}$. The proposed classifier based on logistic regression and its competitors are described below. 
Provided that short bursts or IBIs rarely exist, the output of the classifiers were also smoothed to improve the performance of detections. Hence, the output of this block, $\widehat{Y}$, was finally obtained by removing isolated events below a given time in seconds, $t_B$, applying a filtering procedure similar than \cite{Lofhede2008}.

\subsection{Classification based on logistic regression}
Predictive models based on logistic regression has been successfully employed in a variety of biomedical domains \cite{Hajmeer2003, Liao2007, Ripley2007}. Unlike binary classifiers, that are purely dichotomous, LR provides the class probability for one of the two categories.

In logistic regression \cite{Hosmer2004}, the class probability $\pi_i$ is expressed through a function called logit, related to the feature vector $\mathbf{X_i}$ : 
\begin{equation}\label{eq:log}
 logit(\pi_i) =  \ln\left( \frac{\pi_i}{1-\pi_i} \right) = w_0 + \pmb{w} \cdot \mathbf{X_i}.
\end{equation}
where $\pmb{w}=[w_1, .., w_d]$ is the vector of regression coefficients and $w_0$ is the intercept. The inverse of the above expression, called logistic function, is expressed as:
\begin{equation}
logit^{-1}(\pi_i) = \frac{1}{1+e^{-( w_0 +\pmb{w}\mathbf{X_i})}} = g(\mathbf{X_i},\pmb{w}).
\end{equation}
An important characteristic of the logistic function is that it is bounded between 0 and 1, and thus, it can be used directly to estimate the probabilities of the possible outcomes as $P(Y=1|\pmb{w},\mathbf{X_i}) = g(\mathbf{X_i},\pmb{w})$.

Given the pair of features and labels \{$\mathbf{X_i}, Y_i$\}, the learning process aims at finding the best $\pmb{w}$, which is to maximize the conditional probabilities $P(Y_i | \mathbf{X_i},\pmb{w})$ \cite{Hosmer2004}. This can be achieved by the maximum likelihood estimation (MLE) method. We employed the Newton-Raphson's hill-climbing algorithm, an iterative procedure that maximizes the log likelihood function until a convergence criterion (coefficients leading to the most accurate predictions) is reached.

Once the optimal coefficients, $\widehat{\pmb{w}}$, are obtained, class probabilities, $ \widehat{\pi}_i$, are provided by the logistic function. 
The class membership is decided by a cut-off value $c$ such that $f(\widehat{\pi}_i)>c$ assigns the predictive output value, $\widehat{y}$, to Class 1, and $f(\widehat{\pi}_i) \le c$ assigns $\widehat{y}$ to Class 0. Here, we fixed $c$ to 0.5.

\subsection{Alternate multi-feature classifiers}
In this paper, we have also evaluated the detection of bursts using three widely employed multi-feature, supervised classifiers suitable for binary classification problems: Linear discriminant analysis, support vector machines and the K-nearest neighbor technique. They are briefly described below.

\subsubsection{Linear discriminant analysis}
Linear discriminant analysis can be applied to solve two-class classification problems simply and efficiently based on the characteristics of each class (mean, covariance matrix) \cite{Mclachlan2004}. The LDA classifier finds a discriminant function, i.e. the linear combination of the multi-dimensional features that best separates the two classes. This function provides scores for each class, being the highest values associated to more likely classes.

\subsubsection{Support vector machines}
The SVM is a very popular machine learning technique used in a variety of applications~\cite{Scholkopf2002}.
This classifier uses a transformation (kernel) function to project the data into a higher dimensional space, where classes may become linearly separable.
 More versatile than linear kernel functions, we used a Gaussian radial basis function (RBF) to guarantee the existence of a non-linear decision boundary:
\begin{equation}
 K(x_i,x_j)=\exp^{\|x_i-x_j\|^2/\sigma},
\end{equation}
 where $x_i$ and $x_j$ denote two feature vectors and the kernel parameter $\sigma$ is the radius of influence of the learning samples selected as support vectors by the model. The other parameter in SVMs, the weight of the soft margin cost function ($C$) \cite{Cortes1995}, needs to be adjusted for an optimal decision boundary. While small values provides "local" solutions over-fitting the model, high values tend to simplify boundaries an may not provide accurate separations. Both parameters $\sigma$ and $C$ were optimized by the sequential minimal optimization method (SMO) \cite{Fan2005}.

\subsubsection{K-nearest neighbor}
The kNN is a nonparametric and nonlinear classifier based on proximity criteria. Given the training set of features, the algorithm identifies the $k$ closest neighbor vectors to classify a new instance. The class assigned to the new instance is then decided by majority vote, i.e. the class accounting for more neighbors. The value of $k$ was set as the square root of the number of instances \cite{Duda2012}.

\subsection{Detection by thresholding} 
Thresholding is a simple technique that can be employed when one-dimensional feature vectors $\mathbf{X}$ can be partitioned in two disjoint regions (classes) by a threshold $T$. To find $T$, an optimization procedure that  maximizes the agreements with the gold standard is performed. New instances are then classified by a simple rule: if the feature value exceeds $T$, it is labeled as $Class \;1$, otherwise as $Class \; 0$.

We employed the thresholding approach proposed by Palmu et al. \cite{Palmu2010a}. 
Briefly, it consists on first pre-processing EEG by a band-pass filter with cut-off frequencies $F_l$ and $F_u$, respectively.  Next, the feature (given by the NLEO operator) is computed in $W$-second windows so that values over $T$ provided a first classification, corrected in a second instance by eliminating bursts below $t_{B}$. By means of an iterative process, $F_l$, $F_u$, $W$, $T$ and $t_{B}$ were optimized to obtain a maximum agreement with their gold standard. We simplified this procedure by optimizing $T$ and imposing the remaining parameters (see Section \ref{subsec:optim}).

\begin{figure*}[th!] 
   \centering
   \includegraphics[trim = 5mm 0mm 5mm 0mm,clip=true,width=\linewidth]{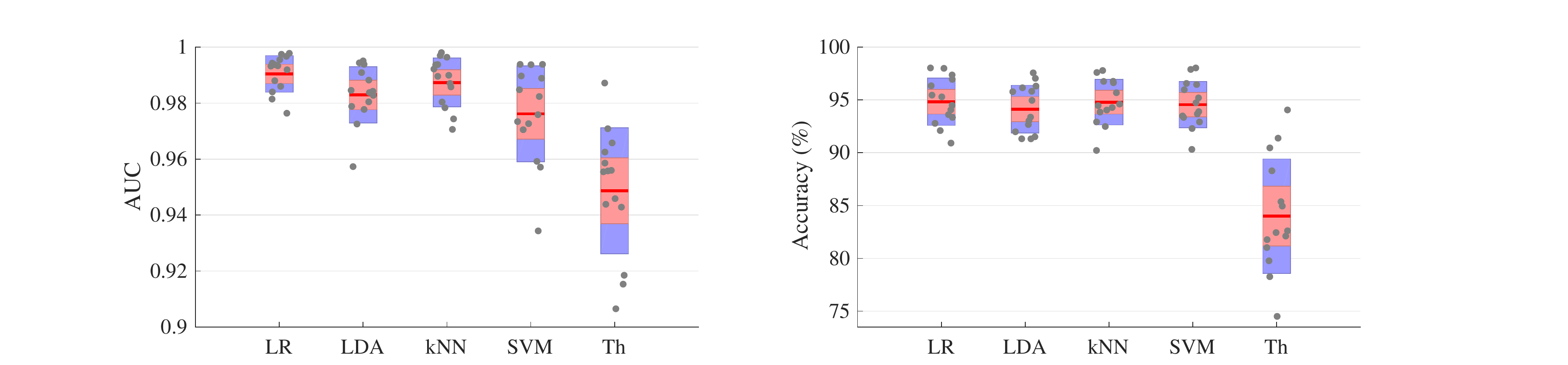} 
   \caption{Performance of all classifiers in terms of areas under ROC curves (AUC, left panel) and  accuracy after performing feature selection. Horizontal red lines denote mean values, red and blue zones represent standard deviations and 95\% confidence intervals (C.I), respectively. Individual values are given by grey circles. Only the performance of $Th$ is significantly below the rest of classifiers (non-overlapping C.I.).}
  \label{fig:box_compar}
\end{figure*} 

\subsection{Measures of agreement and performance}
\label{eval}
To assess the degree of agreement within human raters and between human and automatic classifications, we employed the Cohen's kappa coefficient \cite{Cohen1960}:
\begin{equation}
\kappa = \frac{P_o - P_c}{1-P_c} , 
\end{equation}
where $P_o$ is the observed agreement among raters (the proportion of windows where the observers agreed) and $P_c$ is the probability expected by chance. The upper limit of this statistic ($\kappa=1$) occurs only when there is perfect agreement. The lower limit ($\kappa \le 0$) depends on the marginal distributions and occurs when agreements are  due to chance \cite{Cohen1960}.  

The performance of the classifiers was evaluated by accuracy and receiver operating characteristic (ROC) curves. Accuracy ($Acc$) is defined as the percentage of windows correctly classified over the total number of windows in each labeled EEG. ROC curves represent a sensitivity/specificity pair corresponding to a particular decision threshold. The area under the ROC curves (AUC) summarizes the overall ability of the classifiers to discriminate between the two classes and ranges from 0.5 (random classification) to 1 (perfect classification). 

To obtain unbiased estimations of accuracy and AUC, the performance of the classifiers was examined in infants that did not take part in the training process. Hence, leave-one-out cross-validation (LOOCV) was applied:
\begin{enumerate} 
\item Leave out a single sample from the evaluation subset composed by $N_e$=14 infants.
\item Build the classification model with the remaining data ($N_e$-1).
\item Test the omitted sample with the learned model.
\item Repeat the above steps until each sample has been omitted and tested once.
\end{enumerate}

\section{Results and discussion}
\label{resultat}

\subsection{Setting up automatic detections}
\label{subsec:optim}
Filter cut-off frequencies  $F_l$ and $F_u$ were set to 0.1 -- 30 Hz for multi-feature classifiers. For the simple detection by thresholding, these values were modified (0.5 -- 8 Hz) to meet de requirements of \cite{Palmu2010}. In all cases, features were computed by 75\% overlapping windows of $W$=1 second. This choice is justified by the minimal duration of the bursts in the gold standard but also by a trade-off between reasonable resolution (0.25 s) and computational time. The minimal burst time,  $t_{B}$ was set to 1 second.

We proceeded then to select the most relevant features for LDA, SVM, kNN and LR. 
Feature matrix was composed, per each infant, by 1197 rows (data points) and 8 columns (features). Given that the number of features is low with respect the number of observations, a wrapper feature selection method was employed. The most relevant features were retained by sequential forward selection (SFS), i.e. subsets of features are iteratively combined based on the classifier performance until a maximum is reached.  The maximal performance was evaluated by the mean accuracy yielded by LOOCV. The number of retained features obtained by SFS depended on the classification method. While LDA reached the best accuracy using only two features ($Mm$, $Kt$), SVM needed all excepting $Pw$. For the kNN method, five features were selected ($Mm$, $SD$, $NL$, $AD$, $Hs$) and LR retained three features ($Mm$, $SD$, $NL$). Of note, $Pw$ were discarded by all classifiers, suggesting that it may be redundant or poorly correlated with the labels. On the other hand, $Mm$ constituted the most relevant feature as it was selected in all cases. 

\subsection{Comparison of automatic detections}
\label{subsec:aut}
The performance of the classifiers in terms of AUCs and accuracies are depicted in Fig. \ref{fig:box_compar}). Accuracies were almost identical by using LR, SVM and kNN ($Acc \approx $ 95\%). Little differences exist between these methods when comparing confidence intervals and dispersion. The LDA was slightly below (94\%) and thresholding was the least performing (84\%). LR resulted computationally simpler (uses only 3 features), faster and more intuitive  method than the other classifiers as it provides directly the probability of burst (ease of setting a working point by simply changing the cut-off value $c$).

Our results revealed that thresholding performed poorly compared to burst detection on more immature infants (23 - 28 w PMA), with $Acc$=90\% in average \cite{Palmu2010a}. Indeed, the single feature employed by this algorithm does not describe properly the EEG complexity in older populations of preterm infants. Therefore, additional features need to be included in the classification model.

Even if accuracies provided by LR, SVM and kNN are in the same levels of some of the existing burst suppression detectors in full-terms \cite{Lofhede2010} and above burst/IBI classifiers for preterms \cite{Jennekens2011,Koolen2013}, performances should be compared with caution as they are subject to the design of the gold standard. Thus, the comparison of automatic detections with those obtained by human raters will provide a more realistic idea of the behavior of the classifiers.



\subsection{Visual $vs.$ automatic detections} 
\label{subsec:vis_vs_aut}
The evaluation subset also served to compare the agreements within raters and between raters and the classifiers. For human observations, kappa coefficients were, in average, equal to 0.62. This result improves reported values in populations $<$30 weeks PMA \cite{Murphy2015} (mean $\kappa$=0.58). In terms of accuracy, our mean agreement equals 81\%, a satisfactory result if compared to values obtained in younger cohorts, for instance 81\% in 28 to 30 w PMA infants \cite{Palmu2010a} or 80\% in 29 to 34 w PMA  \cite{Jennekens2011}. In our experiment, discordance was mainly found at the beginning and end of bursts and in few cases concerned a entire burst. 

\begin{figure}[t!] 
   \centering
   \includegraphics[width=\linewidth]{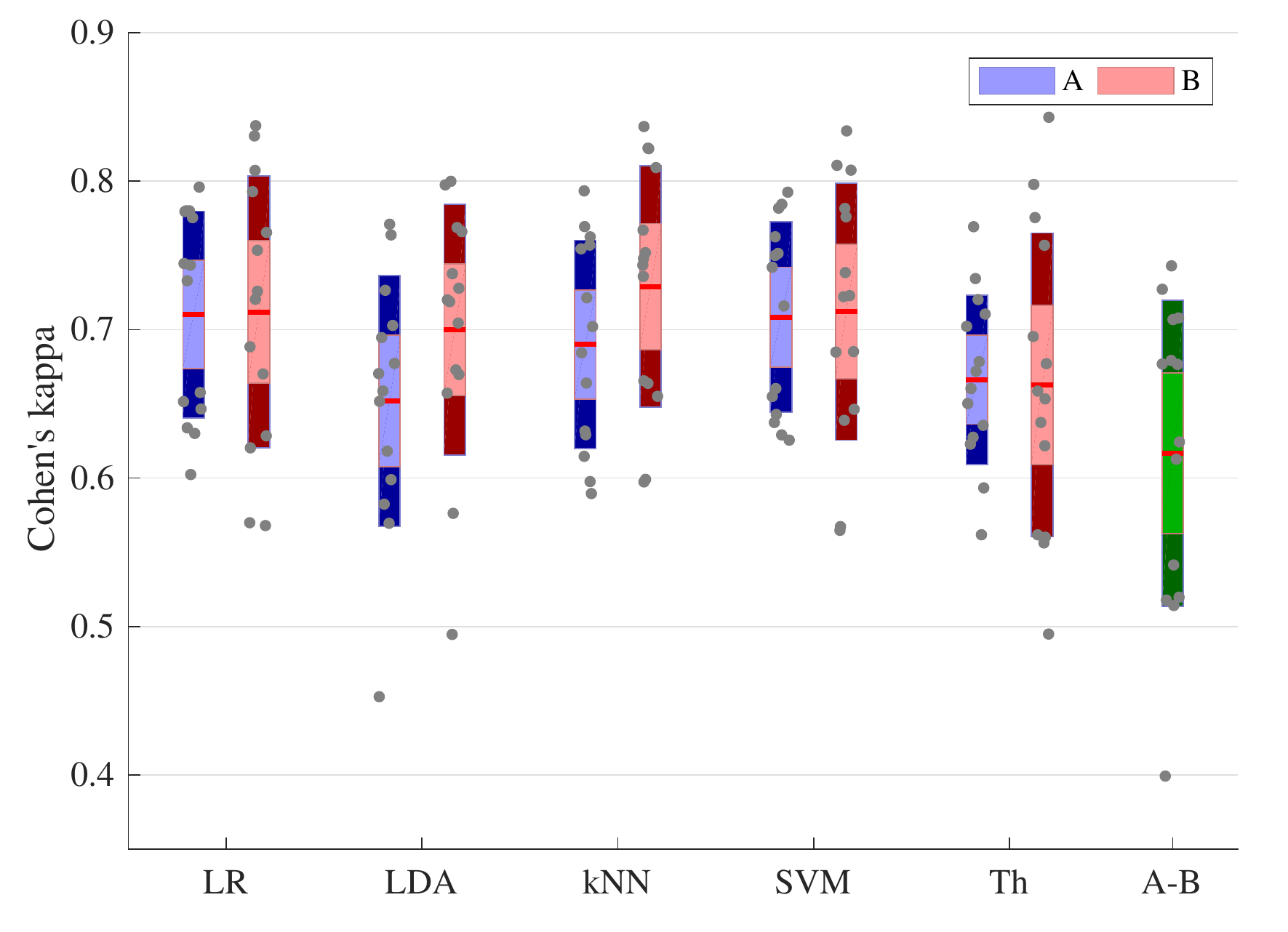} 
   \caption{Comparison of human observations against the automatic detections provided by the five tested classifiers. Blue and red boxes represent kappa coefficients with raters A and B, respectively. None of the A-B pairs showed statistically significant differences in a Mann-Whitney U test. Right green box shows inter-rater agreements. Boxes read as in Fig. \ref{fig:box_compar}.}
   \label{fig:vis_vs_aut}
\end{figure}
 
Comparing the kappa coefficients in Fig. \ref{fig:vis_vs_aut}, it can be stated that automatic-human values are increased with respect to human-human rates. This can be explained by the fact that the gold standard used to train the classifiers is an intermediate reference, i.e. from raters unanimous decisions. Both LR and SVM yielded best averaged human-automatic agreements ($\kappa$=0.71, $Acc$=86\%), but for computational efficiency, LR was our method of choice for the study of maturation presented in Section \ref{appl}.  

\subsubsection{Discontinuity parameters}
\begin{figure}[t!] 
   \centering
   \includegraphics[width=\linewidth]{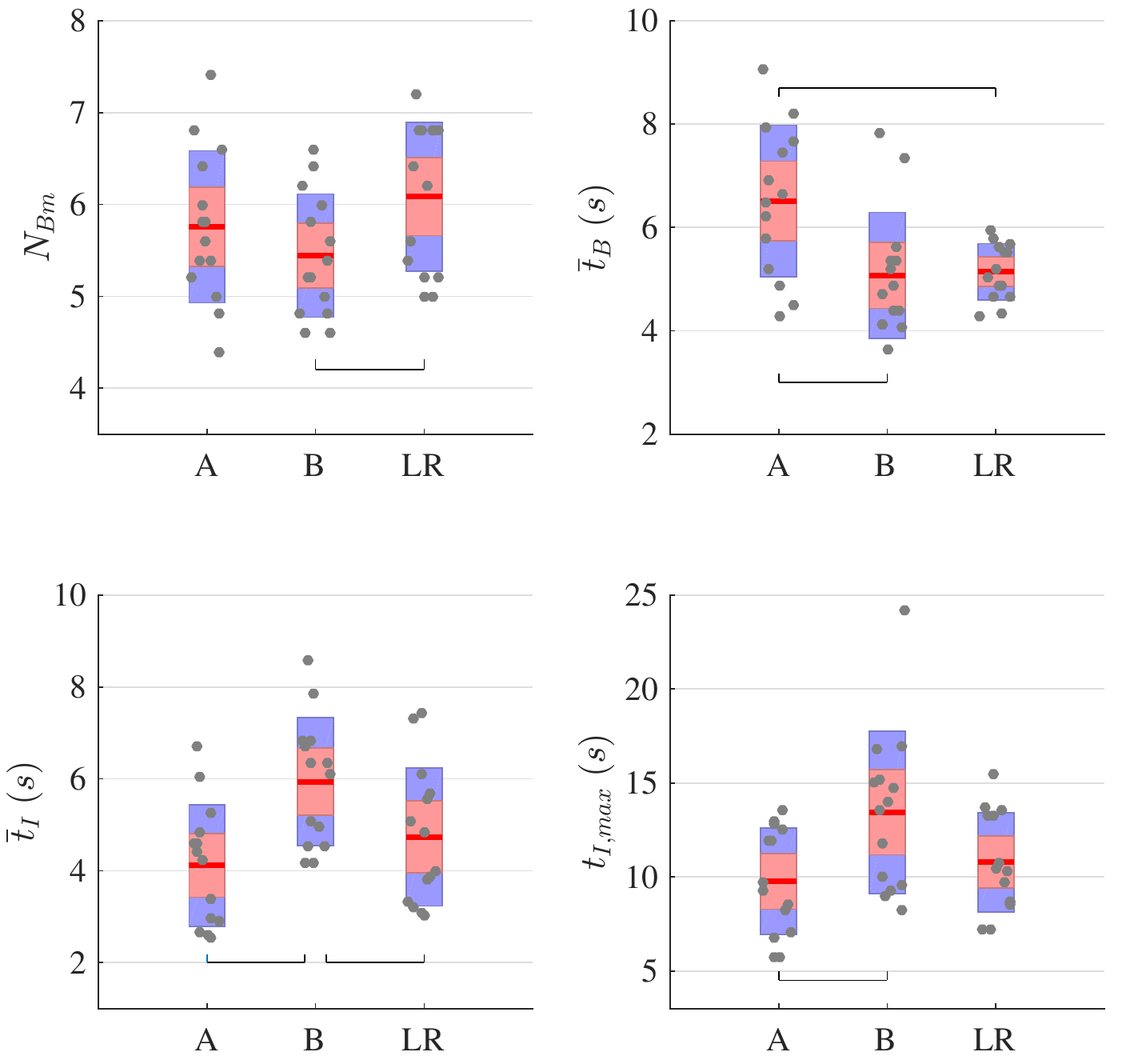} 
   \caption{Characteristics of bursts (discontinuity parameters) according to the raters (A, B) and automatic detection by logistic regression (LR). Horizontal lines grouping a pair of boxes denote statistically significant differences ($p<$0.05) in a Mann-Whitney U test. Interpretation of boxes as in Fig. \ref{fig:box_compar}.}
   \label{fig:charact_bursts}
\end{figure}

In neonatology, maturational patterns are often assessed from the quantitative analysis of EEG bursts. Here, we compared the following measures, also referred to as discontinuity parameters: 
\begin{itemize}
\item Number of bursts per minute ($N_{Bm}$)
\item Mean duration of bursts ($\overline{t}_B$)
\item Mean duration of IBIs ($\overline{t}_{I}$)
\item Maximal duration of IBIs ($t_{I,max}$) 
\end{itemize}
As it can be observed in Fig. \ref{fig:charact_bursts}, values from automated detections are intermediate to those obtained by the raters, excepting $N_{Bm}$ (whose median is over the values obtained by manual marks). Regarding this parameter, differences between LR and B were statistically significant whereas differences between LR and A were not. Significant differences concerning the rest of comparisons with LR cannot be considered relevant as there were also significant differences between A and B (see horizontal lines in Fig. \ref{fig:charact_bursts}). Therefore, automatic detections can be, in general, comparable to human judgment.

\subsection{Study of maturation in a test database}
\label{appl}
We finally computed the above discontinuity parameters in a larger cohort to assess the infants' maturation.  
Infants having sufficiently long periods in quiet sleep ($>$300 seconds) were selected from the main database, discarding too short, unstable sleep patterns. This allowed us to retain $N_t$=20 of 31 newborns who summed up approximately 2 hours of EEG signals. 
Then, we divided the cohort in two set of groups according the median PMA and weight. This allowed to compare the degree of maturity by age (group $G_1^{PMA}$ = [36, 38.2]  versus group $G_2^{PMA}$ = [38.2, 40] w. PMA) and by weight (group $G_1^{W}$ = [1.36, 2.50] versus $G_2^{W}$ = [2.50, 2.86] Kg).  The four discontinuity parameters before described plus the percentage of bursts ($\%_B$) were calculated from the detections yielded by the LR classifier (see Table \ref{tab:burst_matur_QS}).


Significant differences were found in certain parameters regarding weight or age groups. In general, patterns tend to be more continuous as evidenced by the increase of $\overline{t}_B$ and $\%_B$ or the reduction of $t_{I,max}$ in more mature groups. These changes are in concordance with widely accepted maturational criteria, such as the IBI reduction and the prolongation of bursts with increasing PMA \cite{Hahn1989,Biagioni1994}.
Moreover, infants between 35 and 39 weeks PMA rarely exhibit IBIs exceeding 20 seconds, and their mean durations range 4 to 10 seconds depending on the sleep state \cite{Hayakawa2001,Vecchierini2007}, two descriptions that match our results. 


\begin{table}[t!]
\centering
\footnotesize
\begin{tabular}{lcccc}
\toprule
{} &  \multicolumn{2}{c}{PMA (weeks)} & \multicolumn{2}{c}{Weight (Kg)}\\

  {}             					   & 36-38.2 	  			& 38.2-40				&  1.36-2.50		 & 2.50-2.86 \\ 	 
 \midrule
 $N_{Bm}$						& 5.6~$\pm$~0.3 	&  5.9~$\pm$~1.1 	 &  6.0~$\pm$~0.9  &  5.4~$\pm$~0.5\\
 $\overline{t}_B$ (s)  				& 5.5~$\pm$~0.5 	&  6.0~$\pm$~1.5 	 &  5.2~$\pm$~0.7  &  6.4~$\pm$~1.2*\\
 $\overline{t}_{I}$ (s)  					& 5.3~$\pm$~0.8 	&  4.3~$\pm$~0.6* &  4.7~$\pm$~0.8 &  4.6~$\pm$~0.9\\
$t_{I,max}$ (s)  					&14~$\pm$~0.8 	&  11~$\pm$~4.1   &  14~$\pm$~1.8   & 11~$\pm$~3.4*\\
 $\%_B$ 			 		& 52~$\pm$~5.9 	&  57~$\pm$~6.5	 &  52~$\pm$~4.3    &  57~$\pm$~8.1*\\
\bottomrule \\
\end{tabular}
\caption{Discontinuity parameters describing burst activity versus PMA and weight. Asterisks denote statistically significant differences  ($p<0.05$) in a Mann-Whitney U test. }
\label{tab:burst_matur_QS}
\end{table}

\section{Discussion and conclusion}
\label{DC}
The present paper addressed the EEG burst detection problem in very preterm infants who reached term age using multi-feature classification.
The comparative study showed that the classifier based on logistic regression has the best performances in terms of accuracy and computational costs, improving the widely employed thresholding approach and most of the multi-feature classifiers for burst detection proposed in the literature. Indeed, our results benefit from the selection of the channel having the lowest degree of artifacts and from the inclusion of the appropriate burst descriptors by the most relevant features. 

However, the fact that the analyzed EEG corresponded to quiet sleep periods and the high inter-rater agreements might contribute to the high mean accuracy rate (95\%) and mean AUC (0.99) obtained with the best classifier. Moreover, it must be said that these performances do not take into account the possible classification errors in disagreeing zones since the gold standard used to train the algorithms was build from consensual annotations \cite{Palmu2010a}. Hence, accuracies and AUCs could be slightly over-estimated. Nevertheless, bias due to this effect should not concern the choice of the classifier as they influence the performances equally.

Applying the proposed classifier based on logistic regression, we found that parameters describing the discontinuity of bursts by the tested classifiers are in the same range than clinicians' judgments. Therefore, the implementation of this automatic detector would help assessing the infant's maturity in a more repeatable, faster and cost-effective way.


In summary, the main advantage of our proposal relies on its simplicity, reliability and computational efficiency thanks to a logistic regression detector using a single EEG channel. This framework could add new functionality to current bedside monitors, but also it could open the way to home monitors (integrating wearable devices or EEG portable headsets) to follow up maturation in preterm infants after hospital discharge. 


\section*{Acknowledgments}
Authors thank the clinicians from the CHU of Rennes for their involvement in this study. The research for this paper was financially supported by the project INTEM between Rennes University Hospital and the LTSI - INSERM U1099. 

\small
\bibliographystyle{unsrt}

\end{document}